\documentclass{emulateapj}

\newcommand{\mycomment}[1]{}
\newcommand{\UM}{UM\,425}
\newcommand{\UMA}{UM\,425A}
\newcommand{\UMB}{UM\,425B}
\newcommand{\degs}{\ifmmode ^{\circ}\else$^{\circ}$\fi}
\newcommand{\lopt}{\ifmmode L_{2500} \else $~L_{2500}$\fi}
\newcommand{\loglopt}{\ifmmode{\rm log}~L_{2500} \else log$~L_{2500}$\fi}
\newcommand{\logz}{\ifmmode{\rm log}~z \else log$~z$\fi}
\newcommand{\ax}{\ifmmode{\alpha_x} \else $\alpha_x$\fi} 
\newcommand{\aox}{\ifmmode{\alpha_{\rm ox}} \else $\alpha_{\rm ox}$\fi} 
\newcommand{\fcgs}{\ifmmode erg~cm^{-2}~s^{-1[B}\else erg~cm$^{-2}$~s$^{-1}$\fi}
\newcommand{\fnucgs}{\ifmmode {\rm erg~cm}^{-2}~{\rm s}^{-1}~Hz^{-1}\else erg~cm$^{-2}$~s$^{-1}$~Hz$^{-1}$\fi}
\newcommand{\lnucgs}{\ifmmode erg~s^{-1}~Hz^{-1}\else erg~s$^{-1}$~Hz$^{-1}$\fi}
\newcommand{\lcgs}{\ifmmode erg~~s^{-1}\else erg~s$^{-1}$\fi}
\newcommand{\kms}{\ifmmode~{\rm km~s}^{-1}\else ~km~s$^{-1}~$\fi}
\newcommand{\mone}{\ifmmode ^{-1}\else$^{-1}$\fi}
\newcommand{\mtwo}{\ifmmode ^{-2}\else$^{-2}$\fi}
\newcommand{\msun}{\ifmmode {M_{\odot}}\else${M_{\odot}}$\fi}
\newcommand{\lapprox }{{\lower0.8ex\hbox{$\buildrel <\over\sim$}}}
\newcommand{\gapprox }{{\lower0.8ex\hbox{$\buildrel >\over\sim$}}}
\newcommand{\nh}{\ifmmode{\rm N_{H}} \else N$_{H}$\fi}
\newcommand{\nhgal}{\ifmmode{ N_{H}^{Gal}} \else N$_{H}^{Gal}$\fi}
\newcommand{\nhintr}{\ifmmode{ N_{H}^{intr}} \else N$_{H}^{intr}$\fi}
\newcommand{\nhtot}{\ifmmode{ N_{H}^{tot}} \else N$_{H}^{tot}$\fi}
\newcommand{\atoms}{\ifmmode{\rm ~atoms~cm^{-2}} \else ~atoms cm$^{-2}$\fi}
\newcommand{\cmsq}{\ifmmode{\rm ~cm^{-2}} \else cm$^{-2}$\fi}

\newcommand\ha{\ifmmode {\rm H}\alpha \else H$\alpha$\fi}
\newcommand\hb{\ifmmode {\rm H}\beta \else H$\beta$\fi}
\newcommand{\oi}{\ifmmode{\rm [O\,II]} \else [O\,II]\fi}
\newcommand{\oii}{\ifmmode{\rm [O\,II]} \else [O\,II]\fi}
\newcommand{\oiii}{\ifmmode{\rm [O\,III]} \else [O\,III]\fi}
\newcommand{\ew}{\ifmmode{W_{\lambda}} \else $W_{\lambda}$\fi}
\newcommand\whb{\ifmmode{ W_{\lambda}({\rm H}\beta )} \else $W_{\lambda}$(H$\beta$)\fi}
\newcommand\wha{\ifmmode{ W_{\lambda}({\rm H}\alpha )} \else $W_{\lambda}$(H$\alpha$)\fi}
\newcommand{\woii}{\ifmmode{{W_{\lambda}(\rm [O\,II])}} \else $W_{\lambda}$([O\,II])\fi}
\newcommand{\woiii}{\ifmmode{{W_{\lambda}(\rm [O\,III])}} \else
  $W_{\lambda}$([O\,III])\fi}

\shorttitle{Discovery of a Galaxy Cluster Towards UM425}
\shortauthors{Green et al.}

\received{2005 March 1, Accepted May 10}

\begin{document}


\title{Discovery of a Galaxy Cluster 
in the Foreground of the \\ Wide-Separation Quasar Pair UM\,425} 


\author{Paul J. Green,}
\email{pgreen@cfa.harvard.edu}
\affil{Harvard-Smithsonian Center for Astrophysics, 60 Garden Street,
 Cambridge, MA 02138} 

\author{Leopoldo Infante\altaffilmark{1},}
\affil{Departamento de Astronomía y Astrofísica, Pontificia
  Universidad Católica de Chile, Casilla 306, 22 Santiago, Chile} 
\altaffiltext{1}{Visiting Astronomer, European Southern Observatory.}

\author{Sebastian Lopez\altaffilmark{1},}
\affil{Departamento de Astronomía, Universidad de Chile, Casilla 36-D,
  Santiago, Chile}

\author{Thomas L. Aldcroft,}

%\and

\author{Joshua N. Winn\altaffilmark{2},}
\affil{Harvard-Smithsonian Center for Astrophysics, 60 Garden Street,
  Cambridge, MA 02138.}
\altaffiltext{2}{Hubble Fellow}
\vspace{1cm}



\begin{abstract}
We report the discovery of a cluster of galaxies in the field of
UM425, a pair of quasars separated by 6.5\arcsec. Based on this
finding, we revisit the long-standing question of whether this quasar
pair is a binary quasar or a wide-separation lens. Previous work has
shown that both quasars are at $z=1.465$ and show broad absorption
lines. No evidence for a lensing galaxy has been found between the
quasars, but there were two hints of a foreground cluster: diffuse
X-ray emission observed with {\it Chandra}, and an excess of faint
galaxies observed with the {\it Hubble Space Telescope}. Here we show,
via VLT spectroscopy, that there is a spike in the redshift histogram
of galaxies at $z=0.77$. We estimate the chance of finding a random
velocity structure of such significance to be about 5\%, and thereby
interpret the diffuse X-ray emission as originating from $z=0.77$,
rather than the quasar redshift. The mass of the cluster, as estimated
from either the velocity dispersion of the $z=0.77$ galaxies or the
X-ray luminosity of the diffuse emission, would be consistent with the
theoretical mass required for gravitational lensing.  The positional
offset between the X-ray centroid and the expected location of the
mass centroid is $\sim$40\,kpc, which is not too different from
offsets observed in lower redshift clusters.  However, UM425 would be
an unusual gravitational lens, by virtue of the absence of a bright 
primary lensing galaxy.  Unless the mass-to-light ratio of the galaxy
is at least 80 times larger than usual, the lensing hypothesis
requires that the galaxy group or cluster plays a uniquely important
role in producing the observed deflections.

Based on observations performed with the Very Large Telescope at the
European Southern Observatory, Paranal, Chile. 
\end{abstract}




\section{Introduction}
\label{intro}

The probability that a distant quasar is gravitationally lensed by an
intervening potential is sensitive to the volume of the universe, so
lensing statistics place interesting constraints on $\Omega_{\Lambda}$
(\citealt{turner90}, but see also \citealt{keeton02} and references
therein).  Wide separation lensed QSOs ($\Delta\theta > 5\arcsec$)
probe more massive deflectors like groups or clusters of galaxies.
The properties of these dark matter-dominated halos provide strong
tests for the current standard theory of structure formation involving
cold dark matter (CDM). Wide-separation lensed QSOs can measure the
fractional matter density $\Omega_{\Lambda}$ and the rms linear
density fluctuation in spheres of 8$h^{-1}$\,Mpc $\sigma_8$
\citep{lopes04}.  But the detection of wide lensed QSOs has not been
easy.  Only 3 confirmed examples exist with $\Delta\theta > 5\arcsec$:
Q0957+561 (6.3\arcsec\, 
at $z$=1.41; \citealt{walsh79}), RX~J0921+4529 (6.9\arcsec\, at $z$=1.65;
\citealt{munoz01}), and the record-setting quadruple-image quasar
SDSS\,J1004+4112 (14.6\arcsec; \citealt{inada03,oguri04}). 

Wide quasar pairs at similar redshifts are under intense study as
possible wide lenses and have been hunted in large surveys like the
2dF \citep{miller04} and SDSS (\citealt{inada03,fukugita04}).
The lensed quasar candidates are selected to have very similar
redshifts, colors and optical/UV spectra. If no obvious lensing galaxy
is seen between the quasar constituents, deciding whether such pairs
are lensed can be tricky.  Such a case was probed recently by
\citet{faure03} for the 5\arcsec\, pair LBQS\,1429-0053.  Evidence for
lensing may include  
   (1) photometric monitoring and identification of a time delay
between the quasar images, 
   (2) shear in the images of the quasar host galaxy, 
   (3) a weak lensing signal (correlated ellipticity) in galaxies in
the field,
   (4) X-ray diffuse emission and/or 
   (5) an overdensity of galaxies in the field at a redshift
appropriate to a candidate lensing cluster.

Even if these wide pairs are not found to be lensed, they 
can be very illuminating in terms of the study of AGN physics,
the way that twin studies are in medicine.  For instance,
to warrant detailed study of such quasars as candidate lensed
quasars, the pair must have nearly identical spectra, ages and
environment, yet may differ greatly in luminosity.  In this paper,
we study the intriguing pair UM\,425.
 
\section{UM\,425}
\label{um425}

\UM\ is a pair of quasars at redshift $z=1.465$ discovered by
\citet{meylan89} in a search for lenses among 
anomalously bright (presumably magnified) high redshift
quasars. Separated by 6.5\arcsec, the 2 brightest images have nearly
identical optical/UV spectra and close velocities: $\Delta
v_{A-B}=200\pm100$\kms\ from \citet{meylan89}; and $\Delta
v_{A-B}=630\pm130$\kms\ from \citet{michalitsianos97}.  
Both $A$ and $B$ show strong broad absorption lines (BALs),
which occur in a fraction (10 -- 20\%) of optically-selected QSOs
\citep{hewett03,reichard03b}.  Colors of the two images are
indistinguishable from UV through near-IR.  

While UM425 has long been a strong wide lens candidate, 
ground-based optical imaging to $R \sim 24$ reveals no obvious
deflector, arcs, or arclets \citep{courbin95}, whereas a massive lens
should be present to cause the large separation.  

We proposed deep (110~ksec) Chandra observations of the pair, in part
because UM425A (perhaps due to lensing) is one of the brightest known
BALQSOs. X-ray studies of BALQSOs address the debate on whether the
BAL phenomenon is one of orientation, or perhaps accretion rate or
some other intrinsic physical property (e.g., \citealt{becker00,
green01, gallagher02a}). Our analysis of the \UM\, observation (ObsId
3013 obtained on 2001-Dec-13; see \citealt{aldcroft03}, AG03
hereafter) showed that the $\sim$5000 count spectrum of UM425A (the
brighter component) is well-fit with a power law (photon spectral
index $\Gamma=2.0$) partially covered by a hydrogen column of
$3.8\times 10^{22}$~cm$^{-2}$.  This slope is typical for quasars, and
the heavy intrinsic absorption is expected for a BALQSO
\citep{green01}.  Assuming the same $\Gamma$ for the much fainter (30
count) spectrum of UM425B yields an obscuring column 5 times larger.
This X-ray spectral difference (and the difference in $f_X/f_{opt}$
between the two images) could be accounted for in a lens scenario by
differing (or varying) absorbing columns and/or dust-to-gas ratios
along the two sightlines.  Indeed, analysis of the Ly$\alpha$
emission line region in archival HST STIS spectra of the two
components shows - despite their widely disparate $\frac{S}{N}$ - that
both the emission line and absorption profiles differ (AG03).
Spectral differences of this magnitude have been noted previously in
{\em bona fide} lensed quasars (HE~2149-2745 \citealt{burud02a},
SBS~1520+530 - \citealt{burud02b}).

A striking discovery in our Chandra image was significant diffuse
extended emission in the direction of UM425.  Such emission 
arises from the hot gas bound in massive galaxy clusters or groups.
The initial analysis by AG03 suggested that
the cluster $L_X$ (and thereby mass) was probably too low
on its own to explain the wide separation of UM425 in a
lens scenario.  A similar analysis of the X-ray data, informed by a
draft of AG03, was published in a Letter by \citet{mathur03}.
They strongly advocated that the cluster was at the quasar redshift,
while acknowledging that further data were necessary.  

AG03 also analyzed archival HST WFPC2 and NICMOS
images of the field, finding no evidence for a luminous lensing
galaxy.  However, a 3-$\sigma$ excess of faint galaxies in the UM425
field was seen, with plausible magnitudes for a galaxy group
at a redshift ($z\sim0.6$) well-positioned to lens a $z$=1.465
quasar.  The X-ray and optical evidence for a plausible lensing
cluster, as well as the debate over the cluster's redshift,
is what motivated the VLT observations described here.  
For luminosity and distance calculations, we adopt 
a $H_0 = 70$~km\,s$^{-1}$\,Mpc$^{-1}$, $\Omega_{\Lambda} = 0.7$, 
and $\Omega_{M} = 0.3$ cosmology throughout.

\section{VLT Observations}
\label{vlt}

After pre-imaging at the ESO Very Large Telescope (VLT),
we obtained optical spectra of galaxies in the \UM\, field  
in multi-object spectroscopy (MOS) mode.

\subsection{Imaging}

\begin{figure*}
\centerline{
% \resizebox{3.0in}{!}{\includegraphics{A2scaledB2.eps}}
%\plotone{xirvtile.ps}}
\plotone{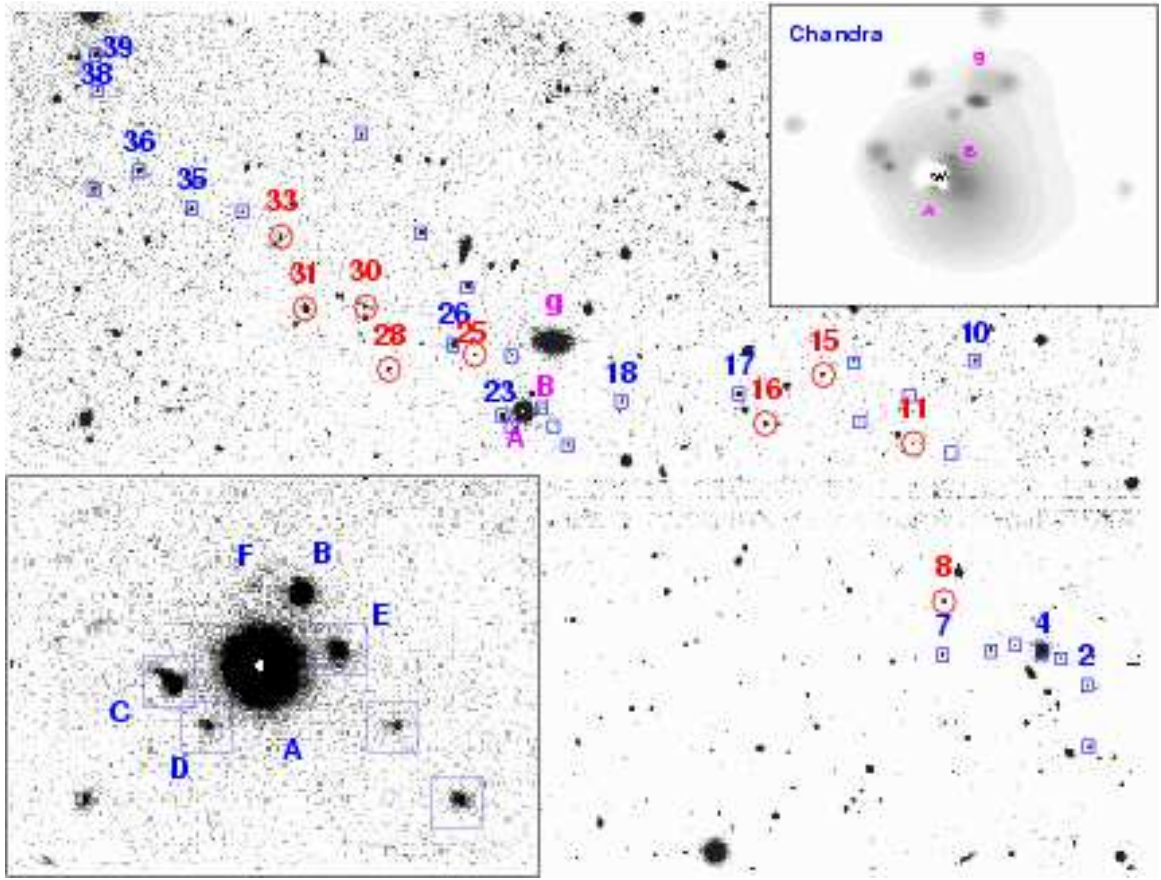}}
\caption{Images of the UM425 field. 
    The FULL FIELD image is in the $R$ band (5\arcmin\, N-S by
6.5\arcmin\, E-W).  Objects whereupon FORS2 slits were placed 
at the VLT are marked. Identifiable spectra were measured for the
numbered objects, whose redshifts are presented in Table~1.  The 9
objects marked with circles have redshifts 0.75 -- 0.78 consistent
with cluster membership as discussed in \S~\ref{spectra}.
   The LOWER LEFT is a 90\arcsec\, N-S and 50\arcsec\, E-W
close-up image in the $I$ band, showing detail near
UM\,425.  The nearby galaxies labeled C-F as in \citet{courbin95}
correspond to 4 of the 6 objects seen on the NICMOS image analyzed
by AG03.  None of their spectra yielded a reliable
VLT redshift.
   At UPPER RIGHT, the {\em Chandra image} inset (1.5\arcmin\, on a
side) is an adaptively smoothed (0.3-3\,keV) image with the central
bright UM425A point source subtracted (adapted from Fig.\,5 of
AG03.
\label{fig_tile}} 
\end{figure*}

We obtained pre-imaging data in Bessel $V$, $R$, and $I$ filters
at the VLT across a 6.7\arcmin\, field at a scale of 0.25$\arcsec$/pixel amid
seeing of 0.8\arcsec\, on  UT~13~March~2004.  Calibration images were
also obtained.  We used SExtractor software \citep{bertin96} to derive
intrumental photometry, using objects detected in the $I$-band (which
had the largest number of detected objects) as a reference catalog in
{\tt ASSOC} mode.  We use {\tt MAG\_AUTO} for $I$ band total
magnitudes, and for colors we use the difference between aperture
magnitudes in apertures of diameter $\sim3\times$FWHM, or about
2.4\arcsec\, (20\,kpc at $z$=0.77; see \S~\ref{spectra}).  Our images
are complete to $V$=24.0, $R$=23.5, and $I$=23.5 (based on a
conservative limit 1\,mag brighter than the turnover in the galaxy
counts histogram).  The wide-field $I$ band image is shown in
Figure~\ref{fig_tile}, along with small-field insets showing the
immediate vicinity of UM425 in optical and X-ray bands.

% Filtername  eff.(nm) FWHM (nm)
% V_BESS+75     554    111.5
% R_SPECIAL+76  655    165.0
% I_BESS+77     768    138.0

% Perhaps more importantly, the fraction of early-type galaxies 
% is spectroscopically only XX\% at redshifts above XX (ref).
% age at redshift 3 was 2.190 Gyr.
% age at redshift 0.77 was 6.993 Gyr.

\begin{figure*}
\centerline{
% \resizebox{3.0in}{!}{\includegraphics{A2scaledB2.eps}}
%\plotone{ivmi.eps}}
\plotone{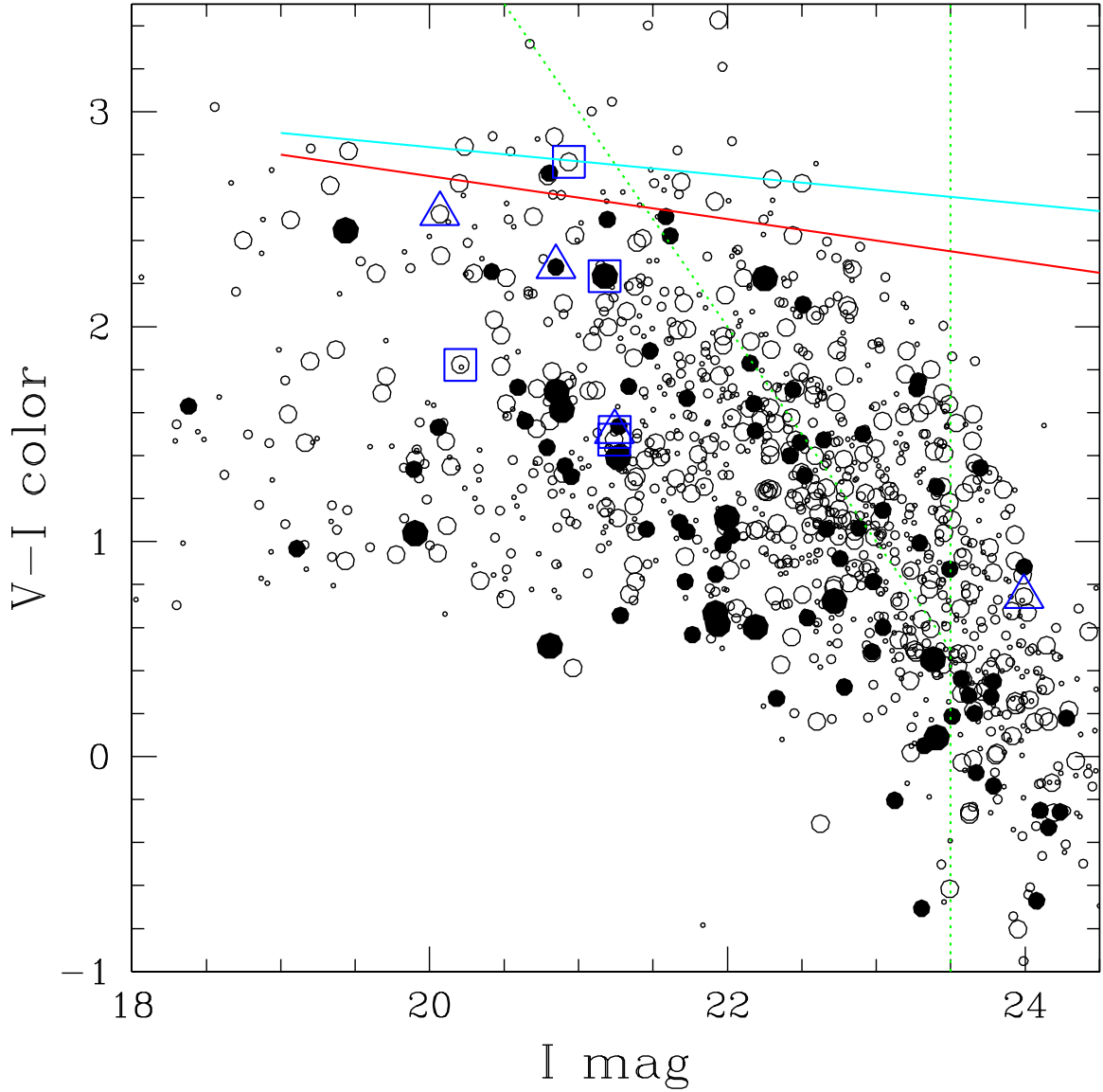}}
\caption{$V-I$ vs. $I$ color-magnitude diagram for objects in the
field of UM425. Objects within 30\arcsec/1\arcmin/2\arcmin\, of UM425
are shown as circles that are large-filled/small-filled/large-open.
The smallest open circles are $>$3\arcmin\, away. 
Galaxies at $z$=0.77 with/without substantial emission lines
are shown as large squares/triangles.  No strong red
sequence is seen in galaxies detected here. The completeness limits of 
$V$=24 and $I$=23.5 are illustrated with dashed lines. 
The lower diagonal line is the expected red sequence locus for this redshift
from DeLucia et al. (2004) that would match the metallicity-luminosity 
relation of Coma.  The solid diagonal line above that is a $z$=0.77
model from Kodama (2004; priv comm).   
\label{fig_ivmi}} 
\end{figure*}

For a cluster at $z$=0.77 (see \S~\ref{spectra}), $R-I$ approximates a 
restframe $U-B$ color, spanning the  4000\AA\, Balmer break. $V-I$
spans an even larger (bluer) range, so is more sensitive to any recent
or on-going star formation.  Figure~\ref{fig_ivmi} shows a $V-I$ vs. $I$
color-magnitude diagram.  
Several recent papers (e.g., \citealt{delucia04}) indicate that a
single-burst model provides a reasonable fit to the red sequence of
high-redshift clusters.  We assume a formation redshift of 3, so that
galaxies at $z$=0.77 are 5\,Gyr old, and $M^*_V\sim$--21
(e.g., \citealt{delapparent03}). Using the {\em
HyperZ} photometric redshift code of \citet{hyperz00}, the expected
colors of bright $M^*$  elliptical galaxies at $z$=0.77
are in the range $V$--$I$$\sim$2.5 -- 2.9, with $21<I<22$.

No strong red sequence is visible among objects within 1\arcmin\, of
UM425.  But our CMD probes only brighter galaxies: $M_V=-20$
corresponds to an observed-frame $I$-band magnitude of about 22.2 at
$z$=0.77.  Many of the cluster elliptical galaxies
composing the sequence are expected to lie fainter or redder than our 
current completeness limits (shown as dashed lines in
Figure~\ref{fig_ivmi}). In addition, the cluster may have a high
fraction of star-forming galaxies, as borne out by our spectroscopy
(\S~\ref{spectra}).  The 
colors of star-forming galaxies, strongly affected by the strength of
the break and \oii\, emission lines, have much larger scatter.  Blue
galaxies like these are candidate progenitors of nearby present-day
faint red sequence ellipticals (\citealt{kodama01,depropris03}).
However, even when galaxies with spectroscopically-identified emission
lines are excluded, substantial ($\sim$0.3mag rms) scatter is
typically seen in cluster red sequence colors.

Deeper imaging in these filters could better determine the number of
galaxies that are likely to be cluster members, and thereby estimate 
an extent and optical centroid for comparison to the X-ray centroid. 
If the diffuse X-ray emission originates in a cluster at the redshift
of UM425, the 4000\AA\, break is close to 1$\mu$m, so that $z$ band or
near-IR imaging is required to estimate cluster membership with
reasonable accuracy.  Space-based imaging would allow for a variety of
other important measurements (see \S~\ref{conclude} below).   

\subsection{Spectroscopy}
\label{spectra}

We obtained spectra of the high-redshift galaxy candidates on
the nights of 10, 12, and 20 May 2004, using the Focal Reducer
Spectrograph (FORS2) at the ESO VLT (U4), using
the grism GRIS\_150I with filter OG590 and a slit width of
6 pixels (0.76$\arcsec$). We effectively covered the spectral range
from $\lambda 6000$\AA\, to 1$\mu$m with a spectral scale of
6.86\AA/pixel, yielding a spectral resolution of FWHM$\sim$19\AA.  We
accumulated 3.26h of exposure time with an effective seeing of about
0.75$\arcsec$. Standard reduction (bias subtraction, correction for
flat field variation, cosmic ray elimination) for both the photometry and
spectroscopy data, as well as rebinning to the observed wavelength for
the spectra, was performed independently using both MIDAS (SL) and
IRAF routines (LI).  The spectral extraction pipeline we used runs
under MIDAS, and simultaneously fits the spatial profiles (with a
Gaussian PSF) and  the sky lines (using a Levenberg-Marquardt
algorithm). Pixels are variance-weighted in the fit.  Cosmic rays are
assigned with infinite variances, so they do not contribute.  
% The final flux is derived from the area of the Gaussian, with
% 1$\sigma$ errors result from the covariance matrix.  

Radial velocities were also measured independently. Results are
identical to within the errors, which are 
themselves conservative (typically 5$\times$ the  difference in
redshift estimates).  In Table~\ref{ztab}, we present the average 
of the two measurements, and the resulting random errors.  We also
tabulate (observed frame) emission line equivalent widths for
\oii\,$\lambda3727$ 

\begin{figure*}
% \resizebox{3.0in}{!}{\includegraphics{A2scaledB2.eps}}
%\plotone{zhisto0.05.ps}
\plotone{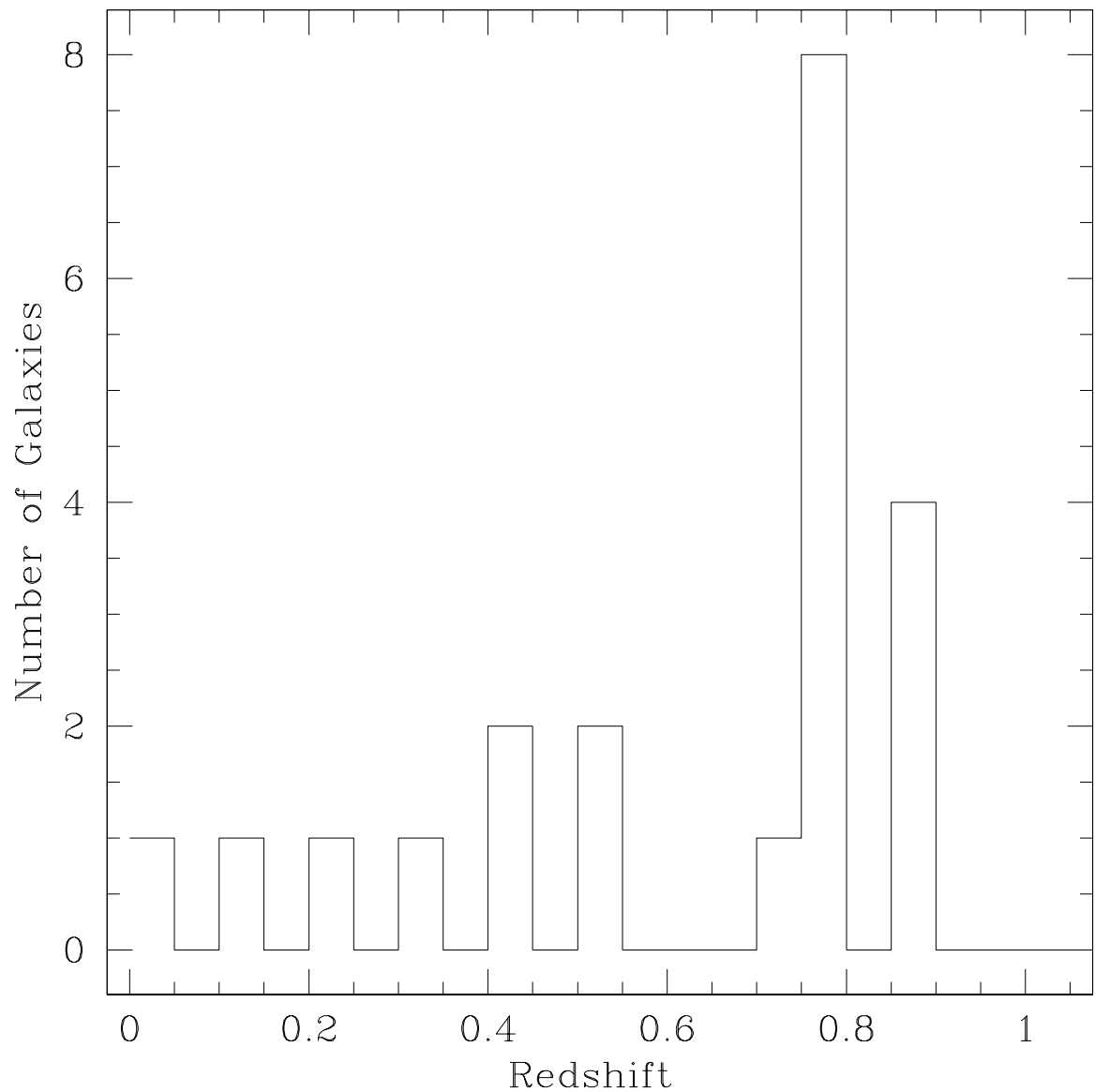}
%\vspace{-2cm}
\caption{Redshift histogram of galaxies in the vicinity of UM425.
Redshifts are accumulated in bins of width $z$=0.05.
\label{fig_zhisto}} 
\end{figure*}

\begin{figure*}
\centerline{
% \resizebox{3.0in}{!}{\includegraphics{A2scaledB2.eps}}
%\plotone{allclustgals0.eps}}
\epsscale{.75}
\plotone{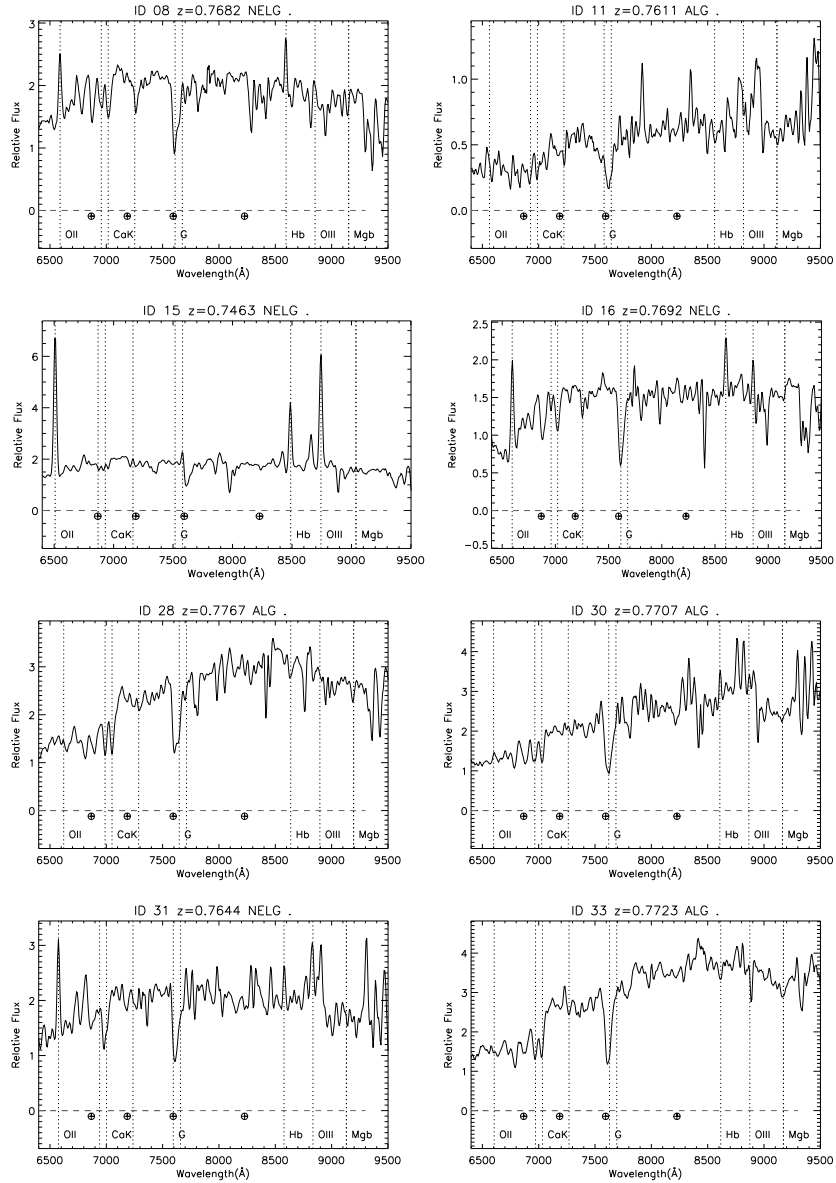}}
\caption{VLT spectra of cluster galaxies.  Flux is marked in units of
  $10^{-18}$\,\fnucgs.  Regions of strong (and sometimes
  poorly-subtracted) sky lines are marked with a symbol near the zero
  flux line.  Prominent intrinsic galaxian absorption/emission lines
  are labeled at the observed wavelength.  Half the galaxies show
  strong emission lines. 
\label{fig_vltspec}} 
\end{figure*}

The redshift histogram in Figure~\ref{fig_zhisto} reveals
a strong cluster of redshifts near $z\sim$ 0.75.  We analyze all 
9 velocities near the cluster redshift (those with 0.6$<z<$0.85),
using the robust biweight estimator of \citet{beers90}, which
includes the velocity errors and yields a redshift  (similar to a
median) of 0.7686$\pm$0.003. The final flux-calibrated spectra for  
cluster galaxies are displayed in Figure~\ref{fig_vltspec}. 
Half the spectra show strong emission line signatures of star
formation. The widest spectroscopically confirmed cluster galaxies 
in the field  (\#8 and 33 in Figure~\ref{fig_vltspec} and
Table~\ref{ztab}) span 2\,Mpc projected separation (at the
cosmological angular scale of 7.4\,kpc/arcsec).

Redshift structures in a given patch of sky trace large scale
structures in the Universe, which may be found in any direction.
What are the chances that a redshift spike such as we find
is unrelated to the diffuse X-ray cluster emission?  We cannot
directly answer this question, but we can estimate the chances of
randomly finding a redshift spike like this in a similar-sized
region on the sky. 

Models of large scale structure, including galaxy luminosity functions
and galaxy evolution in the field, in filaments, and in clusters,
might allow simulations of a patch of sky and hence a test for spikes.
But observations of significant samples of galaxies at these
faint magnitudes are both rare and recent, so that such models are as yet
poorly constrained.  Instead, we turn directly to the best extant deep
wide-field spectroscopic data, in particular the Great Observatories
Origins Deep Survey (GOODS)-North field.   
%         /data/green/prop/AXAF/bals/2001/UM425/paper2/photozs/goods-n-all.dat
There are 1813 reliable spectroscopic galaxy redshifts tabulated by 
\citet{wirth04} and \citet{cowie04} in the GOODS-North field, which
spans about 17\arcmin$\times$10\arcmin.  To test whether a redshift 
spike as strong as the one we find is expected in a random sky
direction, we restrict their full catalog to objects with GOODS
$i$ magnitude $<$23, resulting in a bright subcatalog of 630 redshifts
across the GOODS spectroscopic field.  Due to observing constraints,
the VLT spectra we obtained were constrained to a thin strip on the
sky of approximately  7.5\arcmin$\times$0.6\arcmin
(Figure~\ref{fig_tile}).  We sampled the bright
subcatalog in strips of this size, scanning across the full GOODS
field with a range of step sizes and initial strip positions.  At each
position, we tested  whether at least 21 redshifts existed in a thin
strip. If fewer than 21 were found, we moved another step.  If 21 or
more  were found, we randomly chose 21 within each thin strip. We 
accumulated a histogram of redshifts within each thin strip, using
the same bin width as in our UM425 histogram ($\Delta z$=0.05) and
then we counted the number of redshift spikes of 9 or more galaxies.
For 17,000 such thin sample trials, we found a redshift spike of 9 or
more galaxies in just 3.8\% of samples.  Since the median $I$ mag
of our VLT redshifts is about 21, we tried using a variety of
magnitude-limited subsamples from the GOODS-North sample, and
found that an average fraction of 6.6\% of 21-member
thin subsamples had spikes of 9 or larger.  Since
we {\em only} retained thin subsamples with 21 or more
redshifts, we have biased the subsamples toward significant
structures already, so we consider either of our quoted
fractions to be conservative.  We therefore consider
it very unlikely that this redshift spike is a coincidence
unrelated to the diffuse X-ray cluster emission.

\section{Cluster Properties}
\label{cluster}

\subsection{Velocity Dispersion}
\label{vdisp}

We calculate the galaxy radial velocities as
   $$V_r = \frac{(1+z)^2 - 1}{(1+z)^2 + 1}~c $$
and measure their RMS dispersion about the mean.
Assuming a normal distribution yields $\sigma_v = 1130$\kms\,
for the 9 galaxies within $\Delta z$=0.05 of the mean.  A more
conservative measure might remove the farthest velocity outlier as a
potential interloper, \#15 at $z$=0.7464, which yields a mean (and
also median) optical cluster redshift $z$=0.7688 and $\sigma_v =
670$\kms.  The most robust estimates of scale (similar to RMS
dispersion in the Gaussian case) for velocity samples of this size 
are from the gapper or biweight methods \citep{beers90}, which yield
695$\pm$300 and 552$\pm$280\kms, respectively.  With more redshifts,
these various estimates would likely converge.

For lensed images, the image separation depends only on $\sigma_v$  
of the lens and the ratio of the comoving distances\footnote{We
  calculate angular size distances in our cosmological model using the
  {\tt ANGSIZ} code of \citet{kayser97}.} between the lens and the 
source, $D_{LS}$, and the observer and the source, $D_{OS}$. 
In the SIS model for the lensing mass, and using $z$=0.77,
the observed image separation of 6$\arcsec$.5 implies a cluster
velocity dispersion of $\sigma_v=580 \kms$ or more.  The same model
produces a ``minimum flux redshift'' of 0.6 (AG03), only
10--15\% different (in the relevant angular diameter distance) from
the observed mean redshift.  We thus have strong evidence for an
optical cluster of sufficient mass to lens  the quasar pair.  But are
the optical and X-ray characteristics compatible?

% M_V of Milky Way is -21, of SMC is -17 for H0=50
% M_B* = -20.6 from Schechter 1976
% From 2dF Galaxy groups H. J. Martínez, A. Zandivarez, M. E. Merchán
% and M. J. L. Domínguez 2002, MNRAS, 337, 1441  
% M_Bj* = -19.90 +  5 log (h) so for H=70 M_B*= -20.7
% consistent with 2dFGRS LF for field galaxies

\subsection{Diffuse X-ray Analysis}
\label{xcluster}

Because of the bright X-ray point source coincident with UM425A, it is
difficult to estimate the number of diffuse X-ray photons detected in
the vicinity.  After subtracting the point source as well as possible,
AG03 estimated a lower limit of 51$\pm$12 counts in the
cluster, from a region where the excess above background was clearly
visible.  Now with corroborating optical evidence for a cluster,
we have re-analyzed these data more thoroughly, to better
characterize the cluster X-ray emission.  The somewhat complicated
procedure we describe below reflects the difficulty of studying the
$\sim 200$ X-ray counts from the diffuse emission in close
proximity to the $\sim 3000$ quasar counts.

First, in a 0.5-2~keV image, we fit a PSF model to \UMA\, yielding an estimate
of 3448 total counts.  From the full image, we then exclude a 7\,pixel
(3.44\arcsec)\footnote{Based on PSF modeling we expect about 200 counts from
\UMA\ outside a 7 pixel radius.} region centered on \UMA, and fill the
hole at the background level\footnote{Using a source-free annulus from
70--90\arcsec\, radius, we determined the 0.5-2\,keV background level 
to be 0.0195\,counts/pixel (0.047\,counts/arcsec$^2$).} using the {\tt
dmfilth} tool in CIAO3.1.  We smooth the resulting image using {\tt
  csmooth}  (adapted for CIAO from \citealt{ebeling05}) which
adapts the smoothing scale to result in $\geq$2.5$\sigma$ significance
above background.  The task also generates an image of the smoothing
scale (kernel size map).  Since the PSF depends on energy, we then use
our best-fit quasar spectral model (AG03) as input to
ChaRT\footnote{http://asc.harvard.edu/chart/} and
MARX\footnote{http://space.mit.edu/CXC/MARX/} to create a simulated
PSF at the position of \UMA.  We excise from this simulated quasar
image the same 7\,pixel circular region as in the cluster image, and
smooth with the original smoothing scale map.   Now we have an
quasar image that should be an exact model of the effect of \UMA\, in
the excised and smoothed cluster image.   We thus subtract the
two images. Finally, we refill the excised hole in the cluster image
with a constant level determined from the same size region reflected
across the axis of cluster symmetry in that same image. The regions of
\UMB\, and the nearby galaxy (labeled ``g'' in Figure~1) were also
excised and refilled at the background level.\footnote{Due to the
  small size of these regions and low surface brightness of the
  cluster, filling with counts from the cluster region rather than
  background makes no difference ($<$2counts).} 

\begin{figure*}
\centerline{
%\plotone{cluster_psfsub_smooth_img.ps}}
\epsscale{1.0}
\plotone{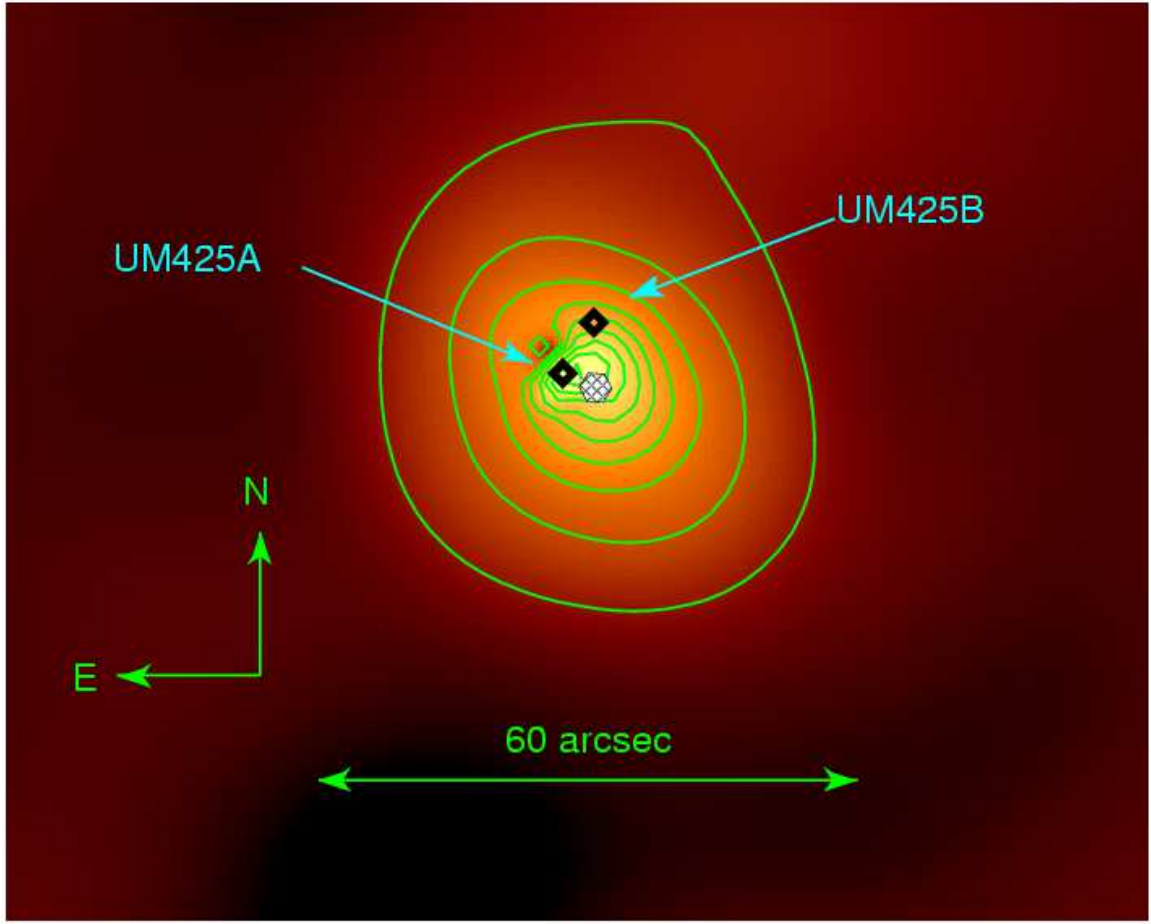}}
\caption{Smoothed 0.5-2\,keV image of the \UM\, region, with QSO
contributions subtracted as described in \S~\ref{xcluster}.
Twelve logarithmic contour levels from 1 to 4 counts/pixel are 
shown.  Positions of the excised QSO images are labeled by black
diamonds.  Estimates for the centroid, ellipticity,
and orientation of the X-ray cluster emission were determined from the
second largest contour.  The white hatched hexagon shows this 
diffuse emission centroid, with a size illustrating its 
uncertainty.
\label{cluster_psfsub_smooth}} 
\end{figure*}

From the smoothed cluster image shown in
Figure~\ref{cluster_psfsub_smooth} it is clear that the peak of
diffuse emission does not coincide with the center of the 
outer contours (which are noticeably elliptical).  This indicates that
the X-ray emitting gas in the cluster is not fully relaxed, and may be
composite. We determine that the cluster center, based on the outer
contours, is at
$11^h\,23^m\,20.5^s\,+01^{\circ}\,37^{\prime}\,46\arcsec$ (J2000).  
The emission peak is approximately 2.6\arcsec\ to the northwest of
that position.  
% Keep it simple and don't mention other techniques that are not described
% The positions of the emission peak counts, from the centroids of
% outer ellipses even using slightly different PSF subtraction and
% smoothing techniques all 
% yield positions within about 3\arcsec\, of this.  
Approximate major and minor axes for the cluster X-ray emission are
($a,b$)=17.7\arcsec,14.3\arcsec\, at position angle $\theta$=40\,deg
(counter-clockwise from North).  A variety of methods (different combinations
of PSF subtraction, smoothing, and the use of ellipse centroids or flux peaks) 
all yield positions within about 4\arcsec.

\begin{figure*}
\centerline{
%\plotone{bothradprof.eps}}
\plotone{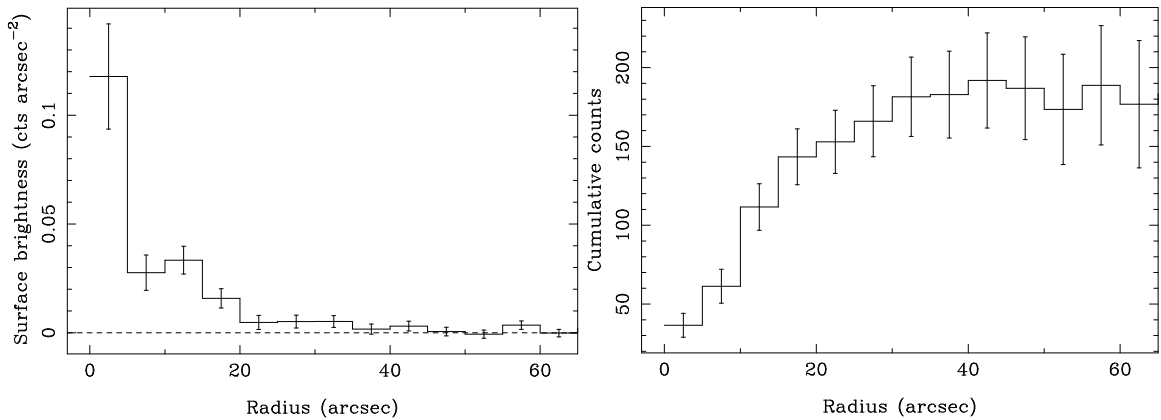}}
%\plottwo{rad_profile.ps}{rad_profile_cumulative.ps}}
\caption{{\em LEFT: } Radial profile (5\arcsec\, annuli) showing
counts per square arcsec in raw 0.5-2\,keV image of the \UM\, region
after subtraction of best-fit 
PSF for \UMA. {\em RIGHT: } Cumulative counts. With this
method, we find a total of 181$\pm$15 counts out to
a radius of 32\arcsec, beyond which the cluster flux is 
not significantly above the background level.
\label{radprof}} 
\end{figure*}

To determine the total flux in the cluster we applied essentially the same
steps described for preparing the smoothed image, but used raw image data with
no smoothing.  This gives our best estimate of the true underlying cluster
emission with all contribution from UM425 (A and B) and the nearby
galaxy removed. 
%
%  on the raw (unsmoothed) image, we fill a 3.5\arcsec\ diameter region
%  centered on \UMA\, (wherein 95.1\% of %total PSF counts should lie)
%  and count (0.5-2keV) photons in 5\arcsec\, annuli until the
%  counts/area reaches ***$1\sigma$ of the background level. 
Figure~\ref{radprof} shows the background-subtracted  radial profile,
as well as the accumulations of counts above background with radius. 
%  We subtract from the cluster total the number of photons
%  (200$\pm$14) that spill out of the 3.5\,pixel annulus excised around
%  \UMA\, 
Based on the point at which the cumulative counts profile flattens out to the
background level, we estimate a total of 181$\pm$25 net
cluster counts within the 32\arcsec\, radial bin\footnote{Neither
  adequate spatial nor spectral information is available from these
  diffuse counts to better constrain the total flux via spectral or
  $\beta$-model spatial fits al\`a \citet{ettori04}}. 

% Boltzmann const k=   8.617342 x 10-5 eV K-1
% logT = log kt(keV) + 7.06

If we assume a Raymond-Smith plasma at $z$=0.77 with a rest-frame temperature 
$kT= 2$~keV and abundance 0.2 solar, the unabsorbed flux is $f = 8.2\pm
1.1 \times 10^{-15}$~erg\,s$^{-1}$\,cm$^{-2}$ (0.5-2~keV).
%  log fx= -14.086
We ignore $K$-corrections, since they are small ($<20\%$) for 
clusters of $T\geq2$keV \citep{jones98}.  Based on the
log$N$-log$S$ of extragalactic diffuse X-ray sources
\citep{boschin02,moretti04}, the likelihood of finding an X-ray
cluster of this brightness (or brighter) within 10\arcsec\, of any
random point on the sky is at most about 4$\times 10^{-4}$.  
Thus this cluster is {\em somehow} associated with UM425, either as
the lensing mass or perhaps (see \S~\ref{hizcluster}) as a host
cluster to two (presumably unlensed) BALQSOs. 

At a redshift of 0.7685 in our adopted cosmology, this spectral model
and flux correspond to an X-ray cluster luminosity of
$L_X$(0.1-2.4~keV)$ = 2.24\times 
10^{43}$~erg\,s$^{-1}$, consistent with a typical cluster. 
This luminosity differs from the earlier estimate of AG03 for two
reasons.  They measured a smaller region, resulting in a {\em lower
limit } on the cluster counts that was a factor 3.5 smaller than our 
more detailed analysis here.  Furthermore, they were forced to assume 
a redshift, so reasonably used the minimum flux redshift of 0.6. The 
revised $L_X$ estimate here is another factor of 1.8 higher, because 
it is now based on the measured optical cluster redshift. 

From \citet{mulchaey98}, who analyzed clusters and
groups together,  both the luminosity and the assumed $T$=2\,keV
temperature correspond roughly to $\sigma_v$=550\kms.\footnote{A similar
estimate of $\sigma_v$=600\kms\, would be found from more recent
samples (e.g., the REFLEX sample; \citealt{ortiz04} after
correcting for different assumed cosmologies).}
This in turn corresponds to a mass estimate sufficient to
achieve the observed splitting of UM425A/B for a singular isothermal
sphere with optimal placement. 
% log Lx= 43.59

% z=0.7685
%   * It is now 13.462 Gyr since the Big Bang.
%    * The age at redshift z was 6.794 Gyr.
%    * The light travel time was 6.668 Gyr.
%    * The comoving radial distance, which goes into Hubble's law, is
%      2700.5 Mpc or 8.808 Gly. 
%    * The comoving volume within redshift z is 82.496 Gpc3.
%    * The angular size distance DA is 1527.0 Mpc or 4.9805 Gly.
%    * This gives a scale of 7.403 kpc/".
%    * The luminosity distance DL is 4775.9 Mpc or 15.577 Gly. 
% 1 Gly = 9.461e26 cm
% 4*pi* ( 15.577*9.461e26)**2 = 2.7293e+57 cm^2, or 57.4368
% for z=0.6 (PaperI) *  luminosity distance DL is 3530.1 Mpc or 11.514 Gly. 
% ratio squared is 1.8306, or 0.2626 in the log

\section{Gravitational Lensing Models}
\label{models}

Even without any quantitative analysis, it is clear that UM~425 would
be an unusual gravitational lens: both the angular
separation and the flux ratio of the quasars are large, and
no primary lensing galaxy has been detected between the two quasars in
deep optical or near-infrared images. Only three out of about 80
well-established lenses have a separation greater than 6\arcsec:
Q~0957+564 (Walsh, Carswell, \& Weymann 1979), RX~J0921+4529
(Mu\~{n}oz et al.\ 2001) and SDSS~J1004+4112 (Inada et al.\ 2003). In
those three cases, there is a central massive lensing galaxy between
the quasar images, whose gravitational deflection is supplemented by a
surrounding galaxy cluster. The large flux ratio between UM~425A and B
($\approx$100 at optical wavelengths, and even larger at X-ray
wavelengths) would be the largest of any known lens. Large
magnification ratios between the two brightest images of a quasar lens
are unexpected, because they generally require the fine-tuned
placement of the source quasar near a caustic of the lensing mass
distribution. In this section, we use the simplest plausible lens
model to illustrate these points.  With a quasar redshift of 1.465, and
assuming a single lens plane at a redshift of 0.77, the critical
surface density for strong lensing is $\Sigma_c = 0.6$~g~cm$^{-2} =
1.6\times 10^{11}$~M$_{\odot}$~arcsec$^{-2}$.

%We use a cosmological model with $H_0=71$~km~s$^{-1}$~Mpc$^{-1}$,
%$\Omega_{\rm M}=0.27$ and $\Omega_{\Lambda}=0.73$. 

For a singular isothermal sphere (SIS: $\rho \propto r^{-2}$), the
fractional cross-section for producing systems with a magnification
ratio greater than $R$ is
\begin{equation}
\frac{\sigma(>R)}{\sigma_{\rm total}} = \frac{4R}{(1+R)^2}
\end{equation}
implying that only 4\% of randomly placed background sources would
produce an image pair with a magnification ratio greater than 100. The
actual probability of finding such a system is even lower because of
magnification bias; systems with large $R$ have a total magnification
of only 2, the smallest possible magnification for a multiple-image
system.

The parameters in the SIS model are the Einstein radius $b$ (which
sets the overall mass scale), the sky coordinates of the center of
mass, and the coordinates and intrinsic flux of the unlensed
quasar. With only the image separation and magnification ratio as
constraints, we have exactly as many parameters as
constraints. Adopting a magnification ratio of 100, the Einstein
radius is $b = 3\farcs 2$ and the center of mass is located only
64~mas from component B, along the A--B line. (The magnification of
each image is proportional to its distance from the lens
center.) The differential time delay between the quasar images is
11~years. To estimate the line-of-sight velocity
dispersion $\sigma_v$, we use the standard relation
\begin{equation}
b = 4\pi \left(\frac{\sigma_v}{c}\right)^2 \left(\frac{D_{\rm LS}}{D_{\rm S}}\right),
\end{equation}
where $D_{\rm LS}$ and $D_{\rm S}$ are the angular-diameter distances
between the lens and source, and between the observer and source,
respectively. The result is $\sigma_v = 540$~km~s$^{-1}$. If this is
interpreted as a single elliptical galaxy of normal mass-to-light
ratio, the corresponding luminosity predicted from the Faber-Jackson
relation is $L/L_{\star} = (\sigma_v/$220~km~s$^{-1})^4 = 40$, which
is ruled out by the non-detection of any galaxy along the A--B line in
optical and near-infrared images. Aldcroft \& Green (2003) ruled out a
galaxy brighter than $0.05L_{\star}$ more than 0\farcs3 from B, and
$0.5L_{\star}$ even if it were coincident with B.

Thus, unless the mass-to-light ratio of the galaxy is at least 80
times larger than usual, the lensing hypothesis requires that a galaxy
group or cluster plays an important role, just as it does for the other 3
known large-separation lenses. There are two indications that there is
indeed a sufficiently massive foreground cluster. First, there is an
overdensity of optically detected galaxies at $z=0.77$, with an
estimated velocity dispersion of 695~km~s$^{-1}$. Second, the diffuse
X-ray emission can be naturally interpreted as a $z=0.77$ cluster with velocity
dispersion of 550~km~s$^{-1}$. The problems are that the centroid of
the optically detected galaxies is very poorly known, and the X-ray
centroid seems to be closer to A than to B (2\arcsec\, vs. 8\arcsec).

One can ask whether it is possible that there is a low-mass and
hitherto-undetected lensing galaxy in an appropriate place between the
quasar images, and that the foreground cluster magnifies the image
separation to its observed large value. A suitably-placed elliptical
galaxy with $L/L_{\star} < 0.5$ would have a velocity dispersion of
$<$185~km~s$^{-1}$, and would need to be supplemented by a cluster
convergence of $\Sigma_c/\Sigma_0 > 0.88$. One issue is that the
domination of the cluster convergence causes the differential time
delay between the quasar images to be small; in this case it is only
1.3~years.  Monitoring by Courbin et al.\ (1995) detected a flare in 
UM425\,B.  The flare is consistent with a microlensing event, but if 
instead was intrinsic to the quasar, the lack of similar burst in
UM425\,A yields a 3-year upper limit to the time delay which is incompatible
with the prediction from the cluster convergence above. If the cluster 
is given a shear of magnitude 0.1 in the direction of the major 
axis of the X-ray isophotes, the time delay is reduced even further 
to 0.8 year. The least massive SIS lens galaxy that produces a time 
delay greater than 3~years has $\sigma=280$~km~s$^{-1}$ and a 
Faber-Jackson luminosity of $L/L_{\star} = 2.7$, which is ruled out.

In short, the interpretation of UM~425 as a lens seems to require 
that at least one of the following possibilities hold:
  [1] The quasar variability observed by Courbin et al.\ (1995) and used to
place an upper bound on the time delay does not reflect intrinsic
variability (due perhaps to microlensing as stated by those authors).
  [2] The observed quasar flux ratio does not reflect the
lensing magnification ratio (due perhaps to an extreme case of
differential extinction or to microlensing). 
  [3] The mass distribution is unprecedently ``dark.'' 
  [4] The X-ray centroid is $\sim$6\arcsec\, (44~kpc at $z=0.77$) from
the true center of mass. 

Relevant to the last possibility, recent measurements of clusters of
galaxies show that all of the following mass centroid estimators may
show significant offsets from each other: 
   [1] X-ray diffuse emission centroid, 
   [2] optical galaxy counts centroid, 
   [3] central dominant (cD) galaxy
position, and [4] weak lensing mass reconstruction.  These offsets may be
attributable to significant substructure, ongoing mergers, or more
generally to the effects of unsettled local dynamical activities on
the intracluster gas.  Chandra and XMM-Newton have shown X-ray cluster
emission centroids offset by 30-70\,kpc from lensing centroid estimates
(e.g., \citealt{belsole05,clowe04, machacek02,jeltema01}). 
For UM425, the cluster X-ray centroid is rather poorly constrained
due to the small number of cluster counts and contamination
by the bright quasar.  Even so, the required dark matter centroid
offset is well within the range seen elsewhere.

% Belsole 2005 A3921 z=.094 17*1.7=28.9
% Clowe 2004 1E\,0657-558 ($z$=0.296)
% machacek 2002 22''*2.96kpc/''=65.12
% jeltema 2001  MS\,1054-0321 at $z$=0.83 40\,kpc betw cD and nearest lump
% um425 7.8*5=39

\section{X-ray Cluster Distance}
\label{hizcluster}

Could the cluster be at the redshift of \UM\, as suggested by
\citet{mathur03}?  If the diffuse emission arises at the quasar
redshift, the X-ray flux for a $T$=3\,keV model (consistent with the
newly-derived $L_X$ below) would be $f = 1.1\pm 0.2 \times
10^{-14}$~erg\,s$^{-1}$\,cm$^{-2}$ (0.5-2~keV), and the luminosity 
would be about 1.5$\times 10^{44}$~erg\,s$^{-1}$.  This is not at all
an unusual cluster luminosity for a massive cluster, so remains an
alternate possibility. 
% log Lx=44.18
The linear scale at that redshift, 8.4\,kpc/arcsec
yields a reasonable cluster size.  A bright cD galaxy should be easily
detected in our $I$ band image.  However, in this high-$z$ scenario,
the quasars are distinct (unlensed) objects.  Either or both are
likely to be hosted by massive galaxies.  Assuming that either of these host
galaxies is a cD progenitor (or if they were to merge as a more
massive cD somewhere along the line connecting them), we again are
faced with significant ($\geq$20\,kpc), but certainly not
unprecedented offsets from the X-ray cluster centroid.

With sufficient X-ray exposure, a definitive redshift can be
measured using the Fe\,K$\alpha$ line from the cluster gas even for 
a distant X-ray cluster (e.g., \citealt{hashimoto04,rosati04}).
Given the low flux of the cluster in the \UM\, field, Chandra exposure
times to achieve an X-ray redshift would be excessive (i.e., weeks),
and an XMM-Newton image would suffer greatly from contamination by the bright
quasar due to XMM's broader PSF.  Worse, BALQSOs themselves may show
strong  Fe\,K$\alpha$ emission \citep{gallagher04}, which would
worsen these contamination issues considerably..

Despite great interest, clusters at redshifts above unity
remain elusive.  Quasars at moderately high redshifts
to date are unfortunately not reliable signposts of massive clusters 
(e.g., \citealt{donahue03}), although there are rare exceptions
\citep{aneta05}.  Even at low redshift, no example of a cluster
with two luminous quasars has been published.  If this is in part the
result of a counter-conspiracy of cosmic evolutions (that quasar space
density peaks near $z$$\sim$2 before massive clusters have formed)
then an X-ray luminous cluster hosting two luminous BALQSOs appears
indeed unlikely. 

Our discovery of an optical cluster renders considerably less probable 
the hypothesis of Mathur \& Williams (2003) is correct that the
cluster is at the  same redshift as UM425.  If there is also a cluster
at $z$=1.465, then very deep spectroscopy, or perhaps deep ($H\sim$22)
multi-band  near-IR imaging could provide redshifts. 

\section{The Binary Interpretation}
\label{binarity}

Mortlock et al. (1999), building on the work of Kochanek, Falco \& Muñoz
(1999), argued that a high degree of quasar spectral similarity is
expected in true binary quasars with sufficient frequency to explain
most of those quasar pairs suspected as lenses but still lacking
detection of a lens galaxy. They further suggest that binary quasars
``are only observable as such 
in the early stages of galactic collisions, after which the quiescent
supermassive black holes orbit in the merger remnant for some time.''
Clearly then, true quasar pairs are of great interest for understanding
the hypothesis that interactions trigger accretion events prior to a
merger (AG03). 

While a redshift of 0.77 for the cluster strengthens the case for the
lens interpretation, we still cannot demonstrate convincingly from the
existing evidence that UM425 is lensed.  Therefore, genuine binarity
for UM425 is not ruled out.  At least two previous quasar pair studies
(\citealt{peng99} for Q1634+267 and \citealt{faure03} for
LBQS\,1429-0053) suffered a similar dilemma.  While the images show
strikingly similar high-S/N spectra, deep imaging revealed no signs of
a lens galaxy, so both studies' judgment weighed toward a binary
interpretation.  However, neither found any evidence for a foreground
cluster as we have here.

\section{Conclusion}
\label{conclude}

UM425A was selected by its anomalous brightness as a lens candidate.
UM425B was found in deep followup to have a similar redshift.
Since these properties were by selection, they are not
by themselves convincing evidence for lensing.  More intriguing
is that both components show evidence for broad absorption lines,
for which the {\em a priori} probability is 1 -- 5\%\, in 
optically-selected quasars \citep{hewett03,reichard03b}.  
Additionally, X-ray and optical evidence for a massive intervening 
cluster presented in this paper are compelling, because at the 
observed significance, the joint probability in the field of 
such a redshift spike ($\sim$7\%) or of diffuse X-ray emission
(0.04\%) is extremely small. 

Based on its large angular separation and flux ratio, and
the absence of a massive lensing galaxy, if UM425 can be confirmed as a lens
it would be an especially interesting one.  There may be a
particularly dark (high $M/L$ ratio) galaxy, or a large offset
between the cluster mass and X-ray emission centroids. Or the
images may suffer extreme differential extinction or variability.

Confirmation of lensing for UM425 is most efficiently achieved with
further deep, ground-based spectroscopy of the field, combined with
deep, high spatial resolution imaging (e.g., using the ACS aboard 
{\em  Hubble}) to  
      (1) better characterize cluster membership via morphology,
magnitude, red sequence colors and photometric redshifts 
     (2) centroid the optical cluster galaxies and luminosity for
comparison to the X-ray centroid and as input to a lensing model,
    and (3) study the field galaxies for evidence of weak
lensing/tangential shear.

PJG and TLA gratefully acknowledge support through NASA contract
NAS8-03060 (CXC).  Work by J.N.W. was supported by NASA through Hubble
Fellowship grant HST-HF-01180.02-A, awarded by the Space Telescope
Science Institute, which is operated by the Association of
Universities for Research in Astronomy, Inc., for NASA, under contract
NAS~5-26555. Imaging and spectroscopy presented
here are based on observations made with ESO telescopes at the Paranal
Observatories under program IDs 271.A-5009(A) and 073.A-0352(B).
Thanks to Tim Beers for providing the {\tt rostat} biweight estimator
code.

{}

\begin{deluxetable}{rllccl}
%\tablewidth{420pt}
\small
\tablenum{1}
\tablecaption{Spectroscopy Results for Galaxies in the UM425 Field}
\label{ztab}
\tablehead{ ID & \multicolumn{2}{c}{RA ~ (J2000) ~ Dec} &  
Redshift & Error\tablenotemark{a}  & $W_{\lambda}$\tablenotemark{b} (\AA) }
\startdata

  2 & 11 23  7.8 & +01 36 19.0 & 0.8646 & 60 &  7.1 \oii  \\ 
  4 & 11 23  8.8 & +01 36 30.1 & 0.0309 & 46 &  21  \ha  \\ 
  7 & 11 23 11.1 & +01 36 28.8 & 0.4201 & 56 &   \\ 
  8 & 11 23 11.1 & +01 36 46.0 & 0.7682 & 44 &  21.1 \oii, 7.1 \hb, 15.1 \oiii  \\ 
 10 & 11 23 10.4 & +01 38  3.6 & 0.4114 & 82 &  \\ 
 11 & 11 23 11.8 & +01 37 36.9 & 0.7611 & 96 &  \\ 
 15 & 11 23 13.9 & +01 37 59.1 & 0.7463 & 35 &  135 \oii, 51 \hb, 78 \oiii  \\ 
 16 & 11 23 15.2 & +01 37 43.5 & 0.7692 & 52 &   37 \oii, 14 \hb, 12 \oiii \\ 
 17 & 11 23 15.8 & +01 37 53.0 & 0.5424 & 55 &  \\ 
 18 & 11 23 18.5 & +01 37 50.5 & 0.8735 & 55 &   54 \oii, 17 \hb, 28 \oiii  \\ 
 23 & 11 23 21.3 & +01 37 46.0 & 0.2470 & 45 &   45 \oiii, 86 \ha  \\ 
 25 & 11 23 21.9 & +01 38  5.7 & 0.7663 & 55 &   34 \oii, 11 \hb  \\ 
 26 & 11 23 22.4 & +01 38  8.7 & 0.5442 & 57 &   14 \hb, 22 \oiii  \\ 
 28 & 11 23 23.9 & +01 38  0.9 & 0.7767 & 51 &   \\ 
 30 & 11 23 24.5 & +01 38 21.0 & 0.7707 & 67 &   12 \oii,   \\ 
 31 & 11 23 25.9 & +01 38 20.5 & 0.7644 & 51 &   26 \oii, 10 \hb, 19: \oiii  \\ 
 33 & 11 23 26.4 & +01 38 43.7 & 0.7723 & 62 &   \\ 
 35 & 11 23 28.5 & +01 38 52.8 & 0.1426 & 53 &   44 \ha   \\ 
 36 & 11 23 29.7 & +01 39  4.9 & 0.8654 & 48 &   41 \oii, 17 \hb   \\ 
 38 & 11 23 30.7 & +01 39 30.6 & 0.8661 & 58 &   61 \oii, 22 \hb   \\ 
 39 & 11 23 30.7 & +01 39 42.2 & 0.3298 & 34 &   11 \hb, 8 \oiii,  77 \ha \\ 
\enddata
\tablenotetext{a}{Redshift error $\times 10^4$.}
\tablenotetext{b}{Observed-frame emission line equivalent width in \AA,
and line identification.}
\end{deluxetable}

%%%%%%%%%% Figure: V-I CMD %%%%%%%%%%%%%%%%%
%%%%%%%%%% End figure  %%%%%%%%%%%%%%%%%%%%%%%%

%%%%%%%%%% Figure: VLT spectra %%%%%%%%%%%%%%%%%
%%%%%%%%%% End figure  %%%%%%%%%%%%%%%%%%%%%%%%

\end{document}